\documentclass{aa}
\usepackage{natbib}
\bibpunct{(}{)}{;}{a}{}{,} 
\usepackage{graphicx}
\usepackage[colorlinks=true,     linkcolor=blue, citecolor=blue, filecolor=blue, urlcolor=blue]{hyperref}

\usepackage{amsmath}
\usepackage{gensymb}
\usepackage{multirow}
\usepackage{subcaption}
\usepackage{txfonts}

\newcommand{\Msol}{M_\odot}

\begin{document}

\title{
Blending from binarity in microlensing searches toward the Large Magellanic Cloud
}
\author{
T.~Blaineau\inst{1},
M.~Moniez\inst{1}
}
\institute{
Laboratoire de physique des 2 infinis Ir\`ene Joliot-Curie,
CNRS Universit\'e Paris-Saclay,
B\^at. 100, Facult\'e des sciences, F-91405 Orsay Cedex, France
}

\offprints{M. Moniez,\\ \email{ marc.moniez@ijclab.in2p3.fr}}

\date{Received 22/03/2023, accepted 03/08/2023}
%

\abstract
{
Studies of gravitational microlensing effects require the estimation of their detection efficiency as soon as one wants to quantify the massive compact objects along the line of sight of source targets.
This is particularly important for setting limits on the contribution of massive compact objects to the Galactic halo.
These estimates of detection efficiency must not only account for the blending effects of accidentally superimposed sources in crowded fields, but also for possible mixing of light from stars belonging to multiple gravitationally bound stellar systems.
}
{
Until now, only blending due to accidental alignment of stars had been studied, in particular as a result of high-resolution space images.
 In this paper, we address the impact of unresolved binary sources that are physically gravitationally bound and not accidentally aligned, in the case of microlensing detection efficiencies toward the Large Magellanic Cloud (LMC).
}
{
We used the Gaia catalog of nearby stars to constrain the local binarity rate, which we extrapolated to the distance of the LMC.
Then we estimated an upper limit to the impact of this binarity on the detection efficiency of microlensing effects, as a function of lens mass. 
}
{
We find that a maximum of 6.2\% of microlensing events on LMC sources due to halo lenses heavier than $30 M_{\odot}$ could be affected as a result of the sources belonging to unresolved binary systems.
This number is the maximum fraction of events for which the source is a binary system separated by about one angular Einstein radius or more in a configuration where light-curve distortion could affect the efficiency of some detection algorithms.
For events caused by lighter lenses on LMC sources, our study shows that the chances of blending effects by binary systems is likely to be higher and should be studied in more detail to improve the accuracy of efficiency calculations.
}
{}


\titlerunning{binarity in microlensing searches}
\authorrunning{Blaineau, Moniez}

\maketitle

\section{Introduction}
Objects cataloged in dense fields are frequently composed of several blended sources.
Not accounting for this may distort the statistical conclusions of the microlensing searches because of its impact on the shape of the light-curve \citep{DiStefano1995}, which can have repercussions on detection efficiency \citep{Wozniak1997}.
Some of the consequences of blending on microlensing have been studied by comparing ground-based images with high-resolution deep space images, obtained notably with the Hubble Space Telescope (HST) \citep{HST2002}. These space images allow one to quantify the impact of accidental alignments of sources in the catalogs of the ground-based surveys, due to the high density of the field \citep{Alibert2005, Tisserand_2007, Wyrzykowski_2010}.
However, in the direction of the Magellanic Clouds, a specific component of the blending remains poorly understood, which is a result of the mixing of light from multiple gravitational bound stars being too close together to be resolved, even by space telescopes, at a distance larger than $50\,kpc$.
Their existence is an additional cause of blending, distinct from the mixing caused by coincidental alignments.
In this paper we study the possible consequences, currently poorly known, of the binarity of the Large Magellanic Cloud (LMC) stars on the detectability of the gravitational microlensing effects they may experience. In particular, we study the case of the detection efficiency of microlensing effects due to high-mass ($>30 M_{\odot}$) Galactic halo objects, which have been recently searched in the LMC direction and excluded as a significant component of the hidden mass of the Galaxy \citep{EROSMACHOcombined}.
In section \ref{section:formalism}, we provide an overview of the fundamentals of the gravitational microlensing effect.
In Section \ref{section:blending} we introduce the blending effects and their impact on the detection efficiency. We introduce the case of multiple sources and distinguish between three blending regimes.
In section \ref{section:correlation}, we present our statistical analysis tools and show that we cannot extract statistical information on the LMC binary systems from HST images because of the separation limit.
In section \ref{section:method}, we describe our methodology to estimate upper values of
the binarity rate in the local volume (at a distance of less than $500\, pc$).
We show how the distribution of distances between the components of star pairs in a complete Gaia catalog population allows us to quantify the rate of widely separated double systems in the Galactic plane.
We explain how we extrapolated the local binarity rate down to $50\,au$ separations for stars closer than $500\,pc$ to the Sun in section \ref{section:fit}.
In section \ref{section:impactLMC}, we detail how we estimated a conservative upper limit of the 
impact of binarity on the microlensing detection efficiency toward the LMC as a function of the projected lens Einstein radius. We discuss the validity domain and limitations of our study, and address the question of the dependence of binarity rates on the stellar type in Section \ref{section:discussion}. 
We conclude and summarize our results in Section \ref{section:conclusion}, and propose some prospects for future microlensing surveys.

\section{Overview of microlensing}
\label{section:formalism}
\subsection{Description of a microlensing event}
The gravitational microlensing effect \citep{Paczynski86}, which was first discovered in 1993 \citep{Alcock1993,Aubourg1993,OGLE1993}, occurs when a massive compact object (the lens) passes close enough to the line of sight of a background source and temporarily magnifies its brightness.
Reviews of the microlensing formalism can be found in \citet{Schneider_2006} and \citet{Rahvar_2015}.
When a single point lens of mass $M_L$ located at distance $D_L$ deflects
the light from a point source located at distance $D_S$,
a situation hereafter referred to as Point-Source Point-Lens (PSPL) event,
an observer receives light from two images not separated in the telescopes.
The total magnification $A(t)$ of the apparent luminosity of the source at time $t$ is given by  the following \citep{Paczynski86}:
\begin{equation}
A(t)=\frac{u(t)^2+2}{u(t)\sqrt{u(t)^2+4}}\ ,
\label{Amplification}
\end{equation}
where $u(t)$ is the distance of the lens to the undeflected line of sight, divided by the Einstein radius $r_{\mathrm{E}}$,
\begin{equation}
r_{\mathrm{E}}\! =\!\! \sqrt{\frac{4GM_L}{c^2}D_S x(1\! -\! x)}\!
\simeq\! 4.5\,\mathrm{au}\times\left[\frac{M_L}{\Msol}\right]^{\frac{1}{2}}\!
\left[\frac{D_S}{10\,kpc}\right]^{\frac{1}{2}}\!\!
\frac{\left[x(1\! -\! x)\right]^{\frac{1}{2}}}{0.5}\!. 
\end{equation}
Here $G$ is the Newtonian gravitational constant, and
\begin{equation}
\label{equation:x}
x = D_L/D_S.
\end{equation}
If the point lens has a constant relative transverse velocity $v_T$
with respect to the point source (PSPL rectilinear event)
and $u(t)$ reaches its minimum value $u_0$ (impact parameter)
at instant $t_0$, then
\begin{equation}
\label{equation:u(t)}
    u(t)=\sqrt{u_0^2+(t-t_0)^2/t_{\mathrm{E}}^2},
\end{equation}
where $t_{\mathrm{E}}=r_{\mathrm{E}} /v_T$ is the lensing timescale,
\begin{equation}
t_{\mathrm{E}} \sim
79\ \mathrm{days} \times 
\left[\frac{v_T}{100\, km/s}\right]^{-1}
\left[\frac{M_L}{\Msol}\right]^{\frac{1}{2}}
\left[\frac{D_S}{10\, kpc}\right]^{\frac{1}{2}}
\frac{[x(1-x)]^{\frac{1}{2}}}{0.5}. 
\label{eq.tE}
\end{equation}
In the case of the study of a population of stellar sources within the LMC, and
assuming that independently of their mass the lenses are distributed in a spherical, isotropic, and isothermal halo with the density distribution described in \citet{Griest1991} and with the parameters used in \cite{EROSMACHOcombined} (hereafter called the standard Galactic halo), the average duration of the events due to the lenses of a given mass $M_L$ is $\langle t_E\rangle\sim 63\,days\times \sqrt{M_L/\Msol}$.
\subsection{Optical depth, event rate, and detection efficiency}
The microlensing optical depth $\tau$ is defined as the probability that the line of sight to a source is within one $r_E$ of a lens (corresponding to $A_{max} > 1.34$ according to Eq. (\ref{Amplification})).
The event rate is the number of microlensing events per star per unit time with $u_0<1$.
This rate is related to the optical depth by the following relation: $\Gamma=(2/\pi)\tau/\langle t_E\rangle$ \citep{Griest1991}.
The event rate is estimated from the duration distribution of the observed events, and it must take into account the observational average detection efficiency that depends a priori on the time sampling, the photometric accuracy, the source population, the background noise, and the search algorithm.
The probability of detecting a given event depends on the parameters of this event ($u_0$, $t_0$, $t_E$, and source magnitude), and the efficiency $\epsilon(t_E)$ is defined as the ratio between the following two numbers:
The numerator is the sum of the detection probabilities of all events with a measured duration $t_E$ and a measured $u_0<1$ (regardless of their other parameters), and the denominator is the total number of expected events with true duration $t_E$, true $u_0<1$, and true $t_0$ within the survey duration.
This efficiency is usually calculated by a Monte-Carlo simulation of microlensing events, combined with observed light-curves of stable stars to produce realistic light-curves of microlensing events, and then subjected to the same analysis as the real data.

\begin{figure}[htbp]
\begin{center}
    \includegraphics[width=9cm]{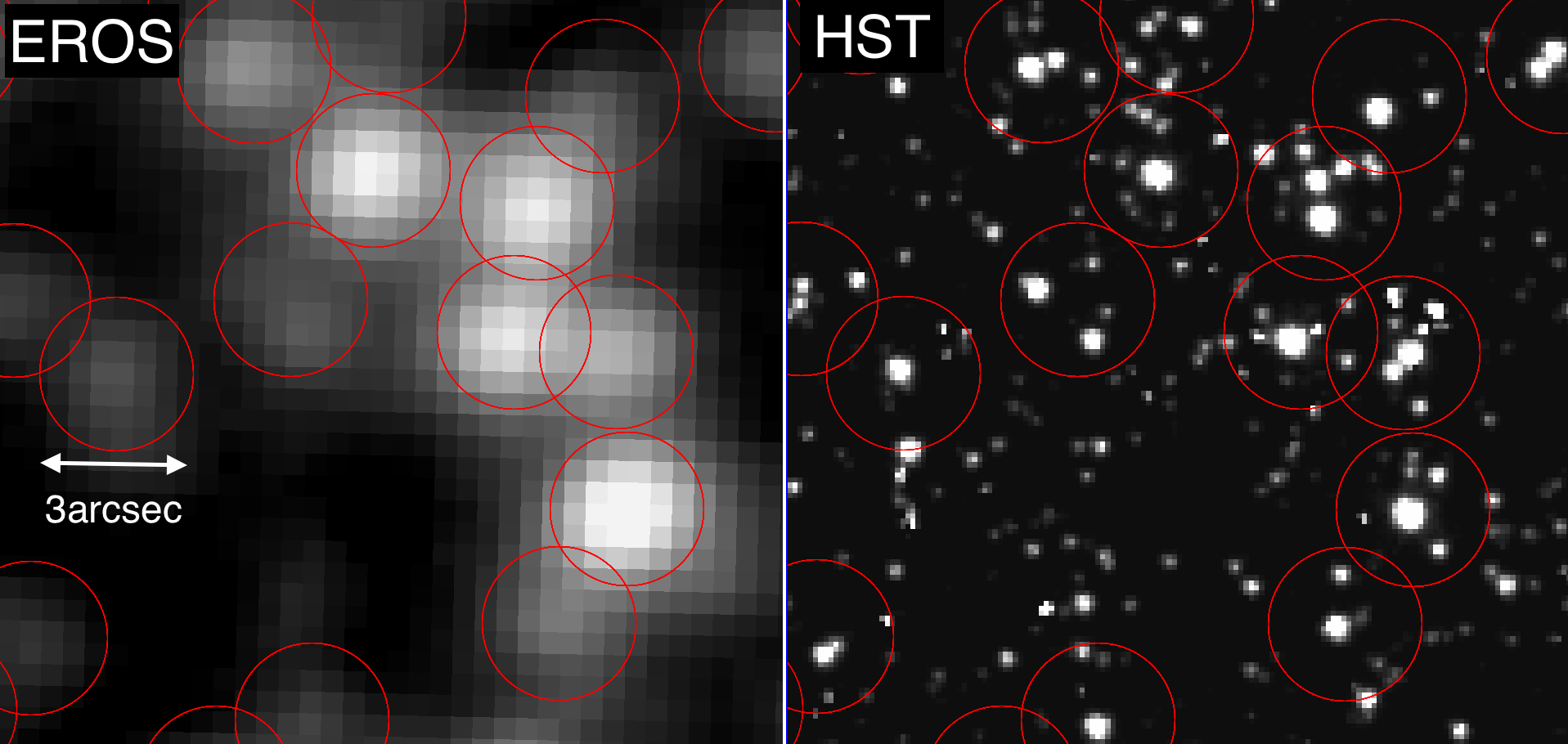}
    \caption[]{
    Images of the same (dense) region of the LMC taken with the EROS camera (left) and the WFPC2 HST camera. The circles are centered on the EROS2 cataloged sources. It is clearly visible that several sources resolved in the HST image can merge into a single EROS cataloged object. We note that even in the HST image, the contribution of hypothetical binary systems is not resolved, and it exceeds the expectation of a random spatial distribution (see Sect. \ref{section:method}).
    }
\label{images-HST-EROS}
\end{center}
\end{figure}

Basic Monte-Carlo efficiency estimates assume that the events are of the PSPL rectilinear type, that is that the magnification curve is given by equations (\ref{Amplification}) and (\ref{equation:u(t)}).
Concerning the studies toward the LMC, it has been established that the contribution of the events that could escape detection due to lens binarity does not exceed $10\%$ \citep{Mroz2019};
the impact of the parallax effects (due to the orbital motion of the Earth around the Sun), which distort the magnification curve, is also negligible toward the LMC \citep{Blaineau2020}.
The subject of this paper is to explore the impact of blending (see Fig. \ref{images-HST-EROS}) on the detection efficiency toward LMC, taking into account not only the blending of accidentally aligned sources already discussed in several works (see for example \citet{Tisserand_2007} and \citet{Wyrzykowski_2011}), but also the -- non-accidental -- contribution of multiple gravitationally bound stellar systems that cannot be directly estimated from LMC spatial images.

\section{Blending effects in microlensing searches: The unknown contribution of binary stars}
\label{section:blending}
In the case where a star, whose flux $\phi_1$ is amplified by a microlensing effect by a factor $A(t)$, is mixed with the light of one or more other stars whose summed flux is $\phi_2$, the evolution of the total flux of the source is the following:
\begin{equation}
     \phi(t) = \phi_1A(t) + \phi_2.
\end{equation}
By defining the blend fraction as the ratio $f_b = \frac{\phi_1}{\phi_1+\phi_2}$,
\begin{equation}\label{eq:blend_flux}
    \phi(t) = (f_b A(t)+(1-f_b))(\phi_1+\phi_2).
\end{equation}
If ignored -- that is using the PSPL approximation --, blending may affect the efficiency of a microlensing search in several ways. On the one hand, the fact that a cataloged object is made up of light from several unresolved stars must be taken into account when estimating the effective number of stars monitored. A microlensing event can indeed occur on each of the stars composing the cataloged source, increasing the number of stars actually susceptible to lensing compared to the number of sources counted in the catalog. On the other hand, the light contribution of other stars modifies the shape of the observed light-curve from a microlensing event, in a
way that changes with the passband.
The efficiency of the search algorithms may then be affected by the following effects:

The shape of the light-curve can no longer be exactly fitted by that of a simple (PSPL) event, which could cause events to be missed with a low-tolerance algorithm
if blending is ignored.
In particular, the apparent or effective magnification $A_\text{eff}$ of the object as observed from the ground is the following:
\begin{equation}\label{eq:blend_A_fb}
    A_\text{eff}(t) = \frac{\phi(t)}{\phi_1+\phi_2} = f_b(A(t)-1) + 1,
\end{equation}
then
\begin{equation}
    \frac{A_\text{eff}(t)-1}{A(t)-1} = f_b < 1,
\end{equation}
showing that the apparent magnification of the observed object is systematically lower than the real magnification $A(t)$ of the lensed star at all times \citep{DiStefano1995}.
Therefore, the measured (or apparent) impact parameter $u_0$ is larger than the physical (true) impact parameter.

As soon as the blend is composed of stars of different colors, the apparent magnification is no longer achromatic: This could reduce the efficiency of a search algorithm that would only consider achromatic events.

The apparent Einstein duration $t_E$ estimated by fitting the light-curve of a blended event with the curve of a PSPL event is systematically shorter than its real value. Since the efficiency of the analysis depends on the measured duration of the event, this means that there is also a modification of the expected number of events as a function of the duration.

All of these effects may impact the number of expected events in a nontrivial way and can potentially modify it, so it is necessary to properly take this blending effect into
account for the interpretation of microlensing effect searches, especially when evaluating the optical depths.
\citet{Mroz_2017} have established that a complete fit of the flux light-curve -- that is to say beyond PSPL approximation --, which takes a blend fraction into account as in Eq. (\ref{eq:blend_flux}), is able to restitute the correct parameters with no bias.
However, the increase in the number of potential sources and the reduced sophistication of detection algorithms mean that detection efficiency can be affected by blending.

As previously mentioned, the confusion of several stars in a single cataloged object has several possible origins: The superposition may be due to an accidental alignment (accidental blending), depending on the field crowding, or it may be due to the multiplicity of the stellar systems (mainly binary blending). The estimation of accidental blending, as it is classically obtained, makes the assumption of a locally uniform density distribution of stars; on the other hand, the impact of superposition due to multiple systems, the binary blending, cannot be estimated in the same way.

Considering the case of two stars blended into a single catalog object, we distinguish three blending regimes depending on their angular separation $\delta$: 
\begin{itemize}
\item
As long as $\delta$ is large compared to both angular Einstein radii of the lens configurations\footnote{If the two stars are not at the same distance $D_L$, then the Einstein rings for microlensing by the same given lens are different.}, we can consider the two stars as independent in the context of the calculation of the expected number of events:
We refer to this mixing regime as classic or ordinary since it is the one that occurs almost exclusively in the case of accidental mixing, and we then observe the addition of a constant flux to an amplified flux.
Previous studies have shown that the impact of this blending regime is small ($<10\%$) and positive on the detection efficiency (\citet{Wyrzykowski_2011,EROSMACHOcombined} and references included).
\item
If the angular separation $\delta$ between the two stars is small compared to the angular Einstein radii of both lens configurations, then both components undergo the same apparent magnification simultaneously, and everything happens as if a single star with a flux equal to the sum of the two stellar fluxes undergoes a PSPL microlensing effect, as in the absence of blending. According to simple geometrical considerations, this regime has a negligible probability of occurring for accidental blending, but not in the case of blending due to binarity. In the latter case, the Einstein radii are identical for the microlensing of both stars. Ignoring this blending regime has no consequence on the optical depth measurement.

\begin{figure}[htbp]
\begin{center}
    \includegraphics[width=9cm]{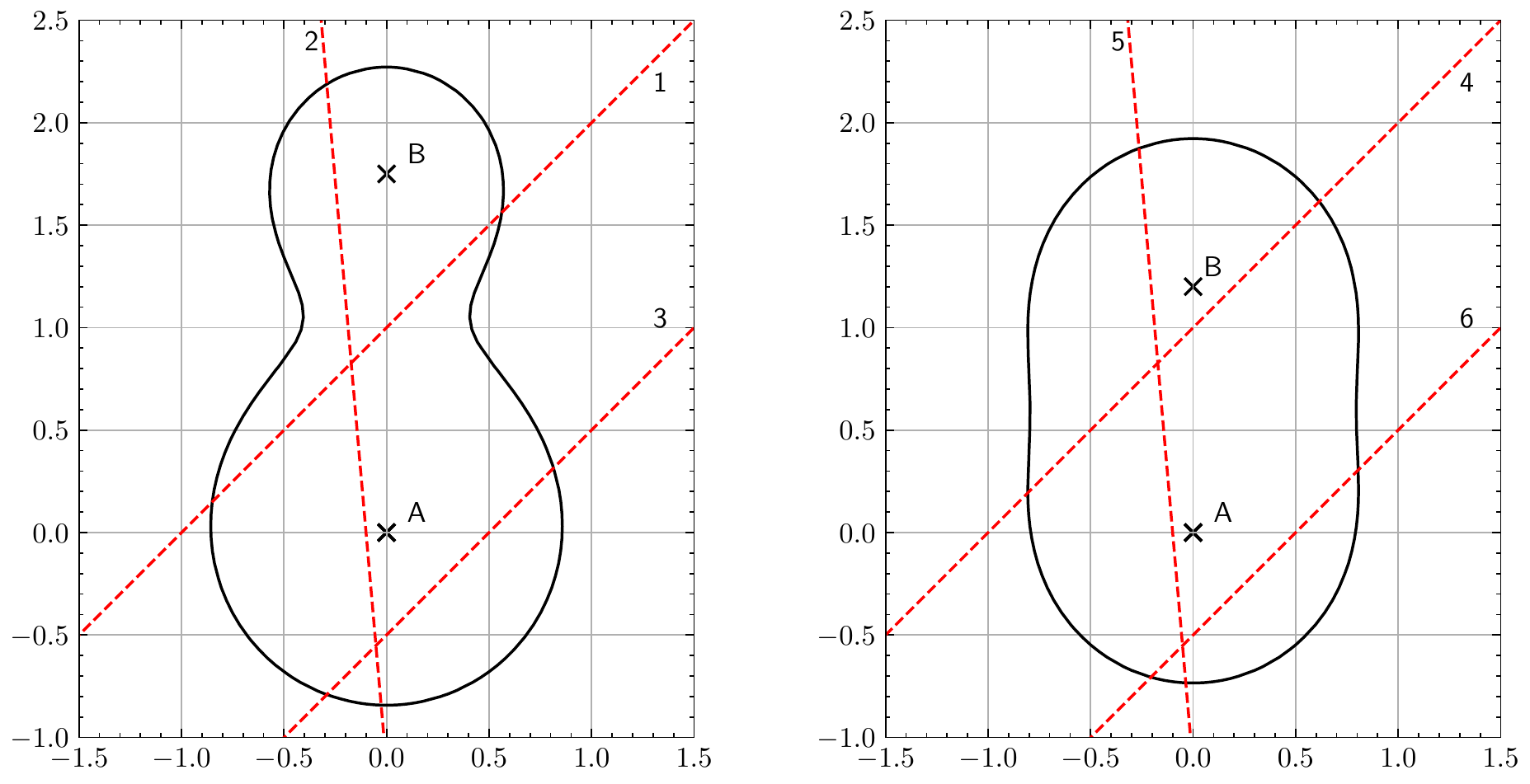}
    \includegraphics[width=9cm]{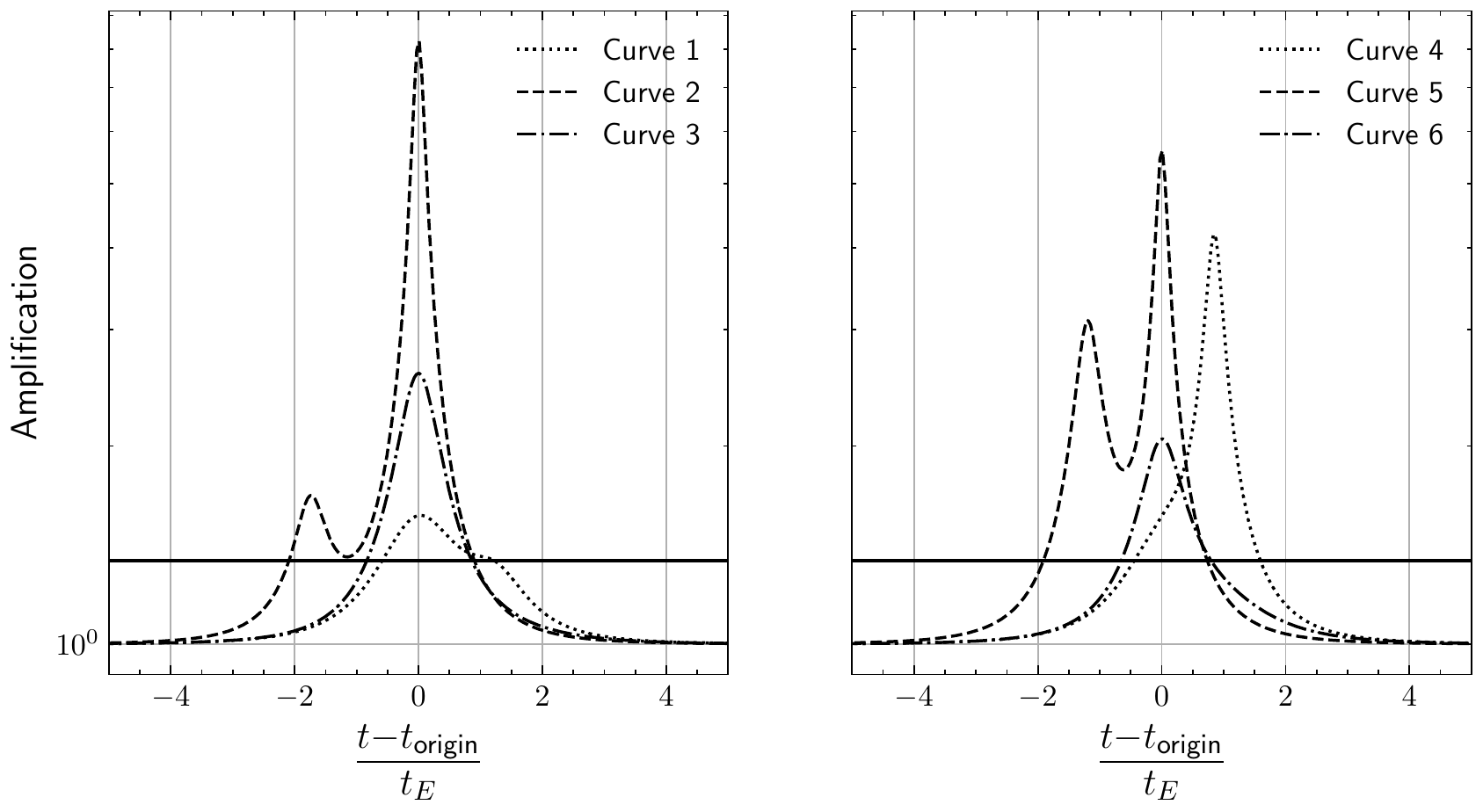}
    \caption[]{
    Lens paths and associated light-curves. \\
    (top) Three different trajectories of a deflector (red dashed) in front of the two stars A and B blended into a single cataloged source; on the left, stars A and B share $80\%$ and $20\%$ of the total luminous flux, respectively, and on the right both stars have the same luminosity. The black outline corresponds to the positions where the deflector must be for the source system to undergo a total magnification of 1.34. The coordinates are in units of Einstein radius, which is the same for both stars (located at the same distance). \\
    (bottom) Magnification curves of the cataloged source system corresponding to the trajectories above, which travelled from left to right. The black line corresponds to a magnification of 1.34. We note that
$t_{origin}$ is the time when the deflector is closest to the star A (and thus does not always coincide with the time of maximum apparent amplification). Curves 3 and 6 are very close to microlensing curves with ordinary blending because the deflector passes far from star B.
    }
\label{trajectories-blend}
\end{center}
\end{figure}
\item
If the angular separation $\delta$ between the two stars is on the order of the angular Einstein radii of both lens configurations, then that would be an intermediate regime, where the light-curve can neither be described by a PSPL microlensing effect nor by the addition of a constant flux to an amplified flux. In this case, we observe the superposition of two events, with different magnifications, and maxima shifted in time. Depending on the geometrical configuration, two clearly separated peaks or a single asymmetric peak may appear. Several examples of magnification curves are shown in Fig. \ref{trajectories-blend}, with the corresponding geometrical configurations of sources and deflectors.
As with the previous regime, this situation can occur only in the case of blending due to binarity, and the Einstein radii are identical for the microlensing of both stars.
Only this regime is likely to cause complications when estimating the average detection efficiency of microlensing effects.
If there were no physical reason for binary systems to exist, this configuration would be extremely rare and could be neglected since the probability of such a small angular distance between uncorrelated objects, even in the densest fields, is negligible.
But when we are dealing with multiple gravitationally bound systems, this situation can happen more frequently.
\end{itemize}
The blending has been studied with the help of HST data,
in two different ways; in the galactic plane direction, some authors (for example \citet{Rahal_2009}) have used the HST images directly to associate objects detected on the ground with stars detected in HST, enabling the content of objects to be studied statistically.
The advantage of this approach is that it benefits from the true spatial distribution of objects, taking any nonuniformity into account, and it is limited only by the resolution of HST.
The other way of using HST data is indirect, and is based on extracting the luminosity function of stars up to 3 magnitudes beyond the detection limit of ground-based surveys. Ground-based telescope images were simulated by injecting stars according to this luminosity function, and the star content of reconstructed objects was then statistically estimated.
Authors using the latter technique assumed that
the spatial distribution of stars is uniform
and neglected the fact that the existence of bound systems results in an excess probability of small separation between components compared to a random distribution \citep{Alibert2005, Smith2007, Tisserand_2007, Wyrzykowski_2010}.
As far as the LMC is concerned, neither approach can take the case of binary systems into account because at the LMC distance ($49.59 \pm 0.09 \pm 0.54\,kpc$ \citep{distance-LMC}) no excess of star pairs can be detected with respect to a random uniform distribution of stars in the HST images, as we show in the next section. 
This means that even the resolution of the space telescopes is insufficient to separate two components of a binary stellar system located at $\sim 50\,kpc$.

To quantify the impact of multiple sources on microlensing toward the LMC, we need a way to count them, based on other data than LMC spatial images.
The following section shows why the limitations of these space-based observations toward the LMC have led us to study a population of the solar environment using Gaia data \citep{2016Gaia}.

\section{Angular and physical distance separation between LMC stars detected by HST}
\label{section:correlation}
The spatial distribution of LMC stars is not expected to be uniform if it includes multiple gravitationally bound systems. This is why
we work with the distribution of the number of pairs of stars as a function of their separation (angular or spatial) and the two-point angular correlation function.
The correlation function is calculated with the Landy-Szalay estimator \citep{Landy1993}, which compares the number of pairs in the data with the number of pairs in a simulation of a uniform spatial distribution in the same spatial domain.
The other way to study multiple systems is to count the number of pairs as a function of their angular separation $\delta$. We expect this distribution to be the sum of a contribution due to fortuitous alignments and an excess at the smallest values of $\delta$ if there are multiple gravitationally bound systems in the set of the pairs. From simple geometrical considerations, the first contribution is a distribution which increases linearly with $\delta$ as long as $\delta$ is small enough so that edge effects do not limit the catalog used. The second contribution, when it exists, is expected for smaller values of $\delta$.

\begin{figure}[htbp]
\begin{center}
    \includegraphics[width=9cm]{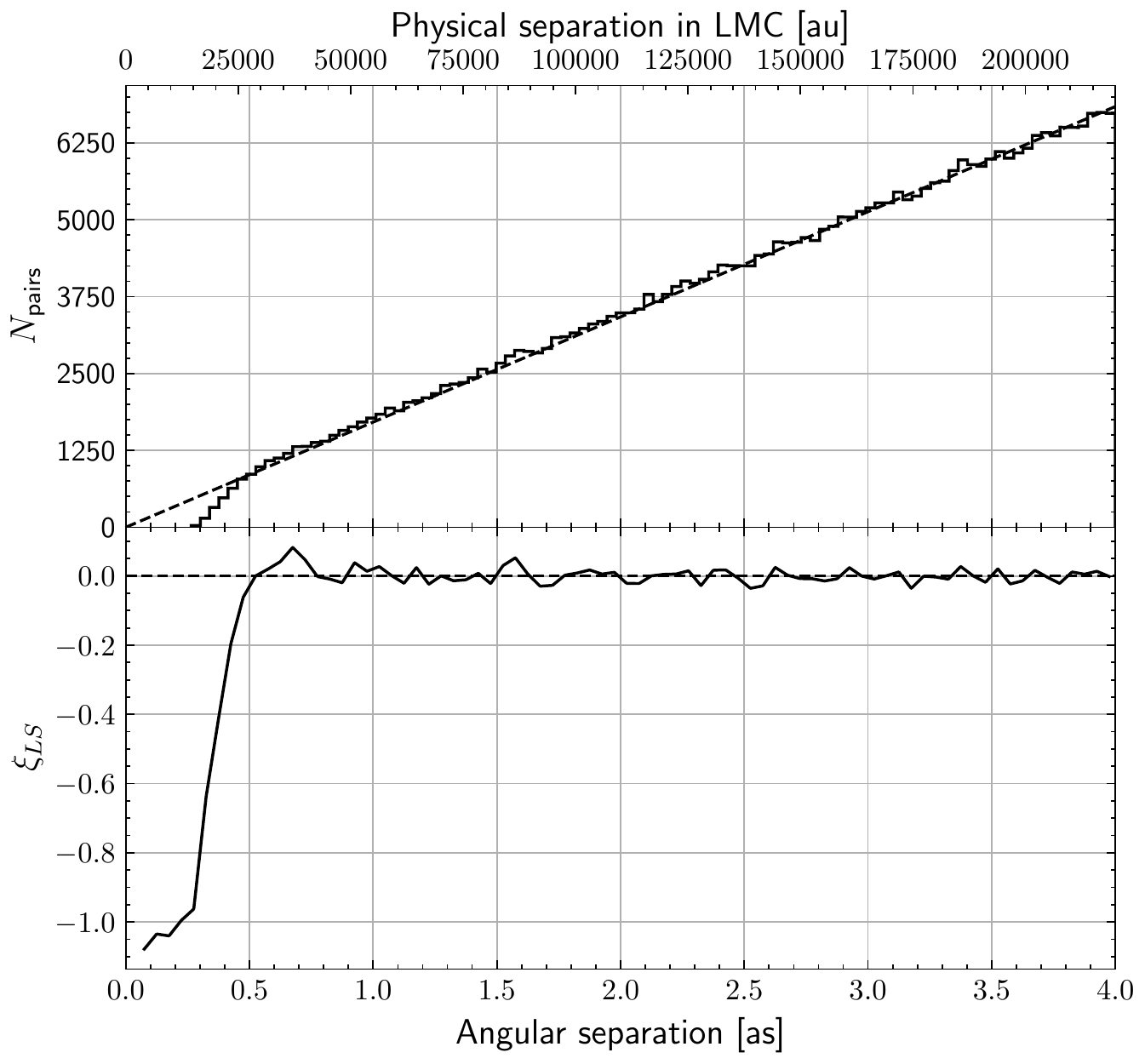}
    \caption[]{
    Number of pairs and two-point correlation function of stars in a HST-crowded LMC field image. \\
    (top) Number of pairs of stars per $1.2''$ interval as a function of their angular separation.
    The dotted line gives the predicted distribution of angular separations in the case of a spatially uniform distribution of stars with the same average star density.

    (bottom) Measured two-point angular correlation function of the detected stars.

    The lower scale is the angular separation $\delta$ and the upper scale gives the corresponding transverse separation in the LMC.
    }
\label{corr_HST}
\end{center}
\end{figure}
Figure \ref{corr_HST} shows the distribution of the number of pairs (top) and the two-point correlation function (bottom), as a function of the angular separation $\delta$ of the stars detected by the SExtractor algorithm \citep{Bertin1996} on an image of a crowded field of the LMC. This image has been obtained by coadding the images taken with the f555w and f814w filters by the Wide Field Planetary Camera 2 (WFPC2) of HST\footnote{HST images hst\_08676\_08\_wfpc2\_f555w\_wf and hst\_08676\_08\_wfpc2\_f814w\_wf, at $(RA,Dec)=(05h06m35.18s,-69\degree 20'45.6'')$.}\citep{HST2002}.
We notice that the correlation function remains zero down to the HST resolution limit (about $0.5''$), decreases below $0.5''$, and is $-1$ below $0.25''$, which indicates that the algorithm can never resolve two stars separated by less than $0.25''$. Since there is no positive correlation for small separations, this shows that we also do not detect an excess of pairs in the HST data over a random distribution beyond a separation of $0.5''$, which corresponds to a transverse separation of $27500\,au$ in the LMC. This situation is not surprising since gravitationally bound systems at such distances are very rare \citep{Dhital_2010,Duchene_2013}. The LMC is therefore too far away for HST to resolve multiple systems that would cause a significant excess of close pairs compared to a random spatial distribution.

\section{Constraining the binarity rate in the LMC}
\label{section:method}
Since the LMC is too far away, we studied a stellar population close to us, using data from the Gaia mission, and then we extrapolated the results to the distance of the LMC.
First, we defined the binarity rate we used and describe our estimation method based on star pair counts.
\subsection{Methodology}
We wished to know the proportion of LMC objects in EROS- or MACHO-type catalogs that are binary systems as a function of their transverse separation $a_t$.
As we have just seen, the LMC is too far away even for space missions to resolve such systems. We therefore studied a population of nearby stars cataloged by Gaia, for which the resolution of systems is possible as soon as $a_t > 500\,au$.
In the following, {\it object} refers to an element of the Gaia catalog (which can be made of a single star or a more complex system) and {\it system} refers to a cluster of objects that we consider to be a gravitationally bound system (typically a resolved binary system).
We then counted the binary stars in the local space, in a situation where they are well separated. 
We defined $n^*_{tot}$ as the total number of objects in the catalog, and
$n^*_{bin}(a_t)$ as the number of objects belonging to a resolved binary system ({i.e.,} made of two cataloged objects), with a transverse separation $>a_t$.
From these simple counts of the resolved double star systems, we could numerically compute the following function:
\begin{equation}
f_{bin}(a_t)=\frac{1}{n^*_{tot}}\frac{dn^*_{bin}(a_t)}{da_t}.
\label{equation:fbin}
\end{equation}
This function is the differential probability 
that a Gaia cataloged object belongs to a double system separated by a projected distance $a_t$. This is a function that we directly derived from the Gaia catalog, as explained in the next sections.
We then defined $F_{bin}(a_t)$, the integrated system binarity rate, as the ratio between the number of binary systems (unlike objects) with
a transverse separation
$>a_t$, to the total number of systems:
\begin{equation}
\label{equation:taux}
F_{bin}(a_t)=\frac{n^*_{bin}(a_t)/2}{ (n^*_{tot}-n^*_{bin})+n^*_{bin}/2}=
\frac{n^*_{bin}(a_t)/n^*_{tot}}{2 - n^*_{bin}/n^*_{tot}}.
\end{equation}
We note that the total number of systems at the denominator corresponds to the sum of the single star systems plus all the resolved binary systems -- not only the systems separated by more than $a_t$--\footnote{As no binary system is resolved in the LMC, we need to use this definition of a rate relative to the systems, instead of a rate relative to the -- unknown -- underlying number of stars.}.
From the definition of $f_{bin}(a_t)$, we obtained
\begin{equation}
\label{equation:proba}
\frac{n^*_{bin}(a_t)}{n^*_{tot}}= \int_{a_t}^{\infty}f_{bin}(a)da,
\end{equation}
which is the fraction of objects belonging to a binary system with a transverse separation $>a_t$.
The number $n^*_{bin}$
is the total number of objects belonging to a binary system resolved by Gaia (depending on Gaia's resolution power).
This number was estimated from
\begin{equation}
\label{equation:bintot}
\frac{n^*_{bin}}{n^*_{tot}}=\int_{a_t^{resol}}^{\infty}f_{bin}(a_t)da_t,
\end{equation}
where $a_t^{resol}$ is the separation limit of Gaia in the catalog we used.
As Gaia's resolution is not a step function, we started integrating from a very small separation $a_t^{resol}=50\,au$, below which we were sure Gaia could not separate two components, but to which we extrapolated our measurements, as shown in Section \ref{section:fit}. This conservative choice then gave us a lower bound on $n^*_{bin}$, and thus an upper bound on $F_{bin}$ in equation (\ref{equation:taux}). This choice is not critical
since we subsequently show that the integral (\ref{equation:bintot}) is much smaller than two, and it therefore has a minor impact on the denominator of Eq. (\ref{equation:taux}).

\subsection{Binarity of stars closer than $500\,pc$ in the Gaia-EDR3 catalog}
\label{section:distances}
We have used the Gaia-EDR3 catalog \citep{2020Gaia_arXiv} for a preliminary assessment of the impact of binarity in microlensing studies. This exploratory work does not aim to extract binarity rates, but rather upper limits on the number of multiple source systems that can give rise to complex gravitational microlensing effects. Our immediate goal is therefore to estimate the maximum fraction of binary systems separated by transverse (or projected) distances $a_t$ comparable to or larger than the Einstein radius of a possible lens projected in the LMC.

In our definitions of the binarity rate, we have neglected the contribution of stellar systems made of more than two stars, since we find only $1\%$ of such systems with an angular size smaller than $10''$ (corresponding to $a_t<6000\,au$ at a $600\,pc$ distance) in the selection of stars defined in the next paragraph.
We therefore neglected this type of system in our treatment of blending effects.
The estimate of the transverse (or projected) physical distance $a_t$ of two stars in the Gaia catalog was deduced from the angular distance $\delta$ between the two components and their annual parallax measurements $\pi$, assuming that they both have an intrinsic uniform and rectilinear motion\footnote{
Since we are trying to quantify binary systems, we must remember that the rotation of the stars of a system around the center of gravity alters the uniform rectilinear motion assumed in the estimation of parallaxes. 
As we are only interested in binary systems separated by more than $50\,au$, with orbital periods on the order of a few centuries, we can consider that the variation of orbital velocity on the apparent trajectory of the stars on the sky has only a negligible impact on the estimation of parallaxes in Gaia.
} \citep{Lindegren2012,Lindegren2021},
by $a_t=1\,au\times\delta/\pi$. With the parallax differences between the components being too imprecise compared to the angular accuracies, we could not carry out a study in three dimensions, which would include the longitudinal distance (along the line of sight). As we explain in the following analysis, parallaxes were just used to convert angular distances into transverse distances and also to reduce the risks of association of very distant objects, accidentally located on the same line of sight, by considering only pairs of stars with compatible parallaxes.

\begin{figure}[htbp]
\begin{center}
    \includegraphics[width=6cm]{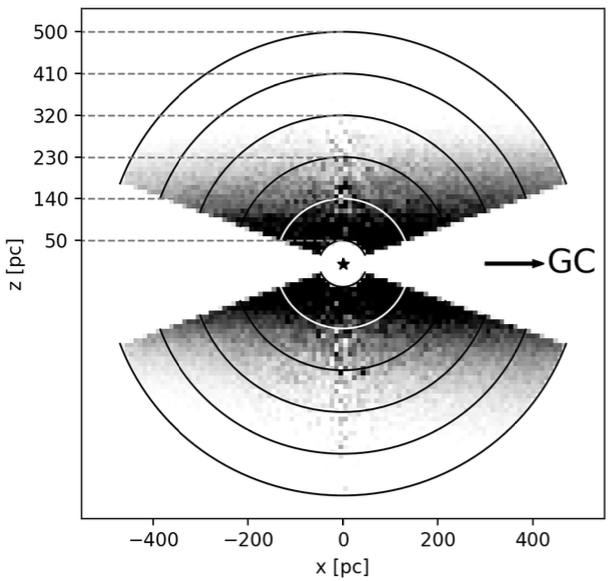}
    \caption[]{
    Projection perpendicular to the Galactic plane of the spatial distribution of the stars of the Gaia catalog, and representation of the limits of the shells used in our study. The representation is centered on the Sun.
    }
\label{shells}
\end{center}
\end{figure}

We selected the stars of the Gaia-EDR3 catalog \citep{2020Gaia_arXiv}, whose apparent magnitude is $3<g<18$, a domain in which the catalog is complete, and thus not biased \citep{2020arXiv_gaia_catalogue_validation}. Then we limited our sample to stars closer than $600\,pc$ (corresponding to a parallax $\pi > 1.66\,mas$). At this distance, a separation of $1''$ corresponds to $600\,au$. We also required that the parallax accuracy be better than $20\%$. The bias \citep{2018Luri} induced by this selection can be neglected because we rejected only $2\%$ of the stars that are the most distant of our sample. Finally, we have restricted our study to Galactic latitudes higher than $20\degree$ to avoid the very crowded and inhomogeneous areas of the Galactic plane (figure \ref{shells}).

\begin{figure}[htbp]
\begin{center}
    \includegraphics[width=9cm]{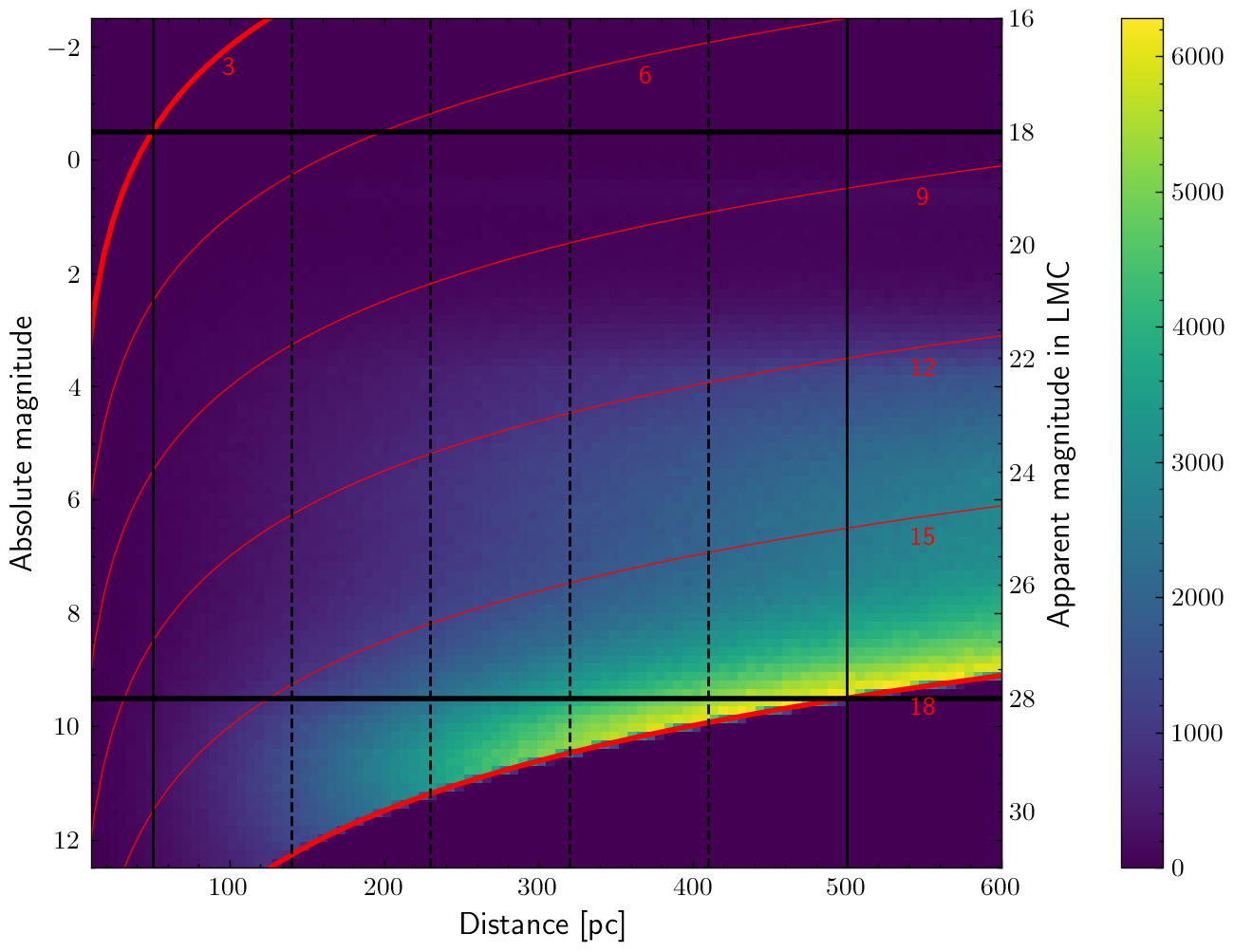}
    \caption[]{
    Distribution of the number of stars in Gaia EDR3 as a function of their distance and absolute magnitude (left scale). The right scale gives the apparent magnitude that these stars would have in the LMC. The Gaia catalog is complete between the thick red curves. The red lines are the iso-apparent-magnitude lines in Gaia in the (distance, absolute magnitude) plane. We restricted the study to stars within the absolute magnitude range delimited by the thick horizontal black lines ($-0.5<G<9.5$), which ensures the homogeneity of the stellar type over the volume we considered; the vertical solid lines correspond to the distance limits of our sample. The vertical dashed black lines delineate the distance domains of the shells we consider in our analysis (between $50$ and $500 \,pc$).
    }
\label{mag-vs-dist}
\end{center}
\end{figure}
Figure \ref{mag-vs-dist} shows the distribution of the absolute magnitude $G$ of the stars of the Gaia catalog closer than $600\,pc$, as a function of their distance, estimated by their parallax. This catalog is complete between the lines of equal apparent magnitude marked by the thick red curves ($g=3$ and $g=18$).
We defined five shells (Fig. \ref{shells} and \ref{mag-vs-dist}), delimited by the spheres of radii (50, 140, 230, 320, 410, and $500\,pc$), and studied the pairs of stars in each shell separately (Table \ref{tab:shellcounts}).
These five shells define a fiducial volume that is not affected by the $600\,pc$ cutoff.

\begin{table}
    \begin{center}
\begin{tabular}{lccccc} \hline \\  [-1ex]
             & $a_t$ for $\delta=2''$ & \multicolumn{2}{c}{number of stars} \\
shell limits & separation & in shell & effective \\
\hline
$[50-140]\,pc$ & $[100-280]\,au$ & 160409 & 158000 \\
\hline
$[140-230]\,pc$ & $[280-460]\,au$ & 481496 & 465000 \\
\hline
$[230-320]\,pc$ & $[460-640]\,au$ & 867259 & 807000 \\
\hline
$[320-410]\,pc$ & $[640-820]\,au$ & 1259719 & 1109000 \\
\hline
$[410-500]\,pc$ & $[820-1000]\,au$ & 1634717 & 1313000 \\
\hline
\end{tabular}
\centering
\caption[]{Limits and contents of the shells (see text).
}
\label{tab:shellcounts}
\end{center}
\end{table}
These subdivisions allowed us to verify that the separation distributions of the binaries are indeed functions of the physical distances $a_t$, independently of the angular distances $\delta$, as long as they are greater than the separation power of Gaia.
To ensure that the distributions studied in the five shells considered are all complete, we only studied the stellar population whose absolute magnitude is in the range $-0.5<G<9.5$ (between the thick black lines in Figure \ref{mag-vs-dist}). Finally, we only considered pairs of stars with a magnitude difference of less than $2.5$, beyond which we could neglect the impact of the less luminous component on the light-curve of a microlensing effect.
\begin{figure}[htbp]
\begin{center}
    \includegraphics[width=9cm]{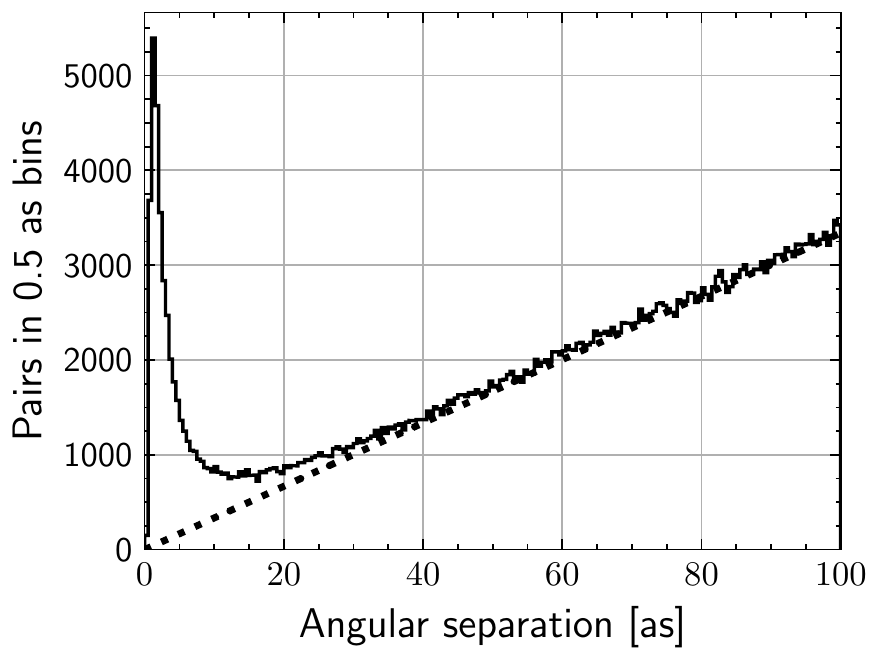}
    \caption[]{
    Number of star pairs per $0.5''$ interval as a function of their angular separation.
    The dotted line gives the predicted distribution in the case of a spatially uniform distribution with the same average star density.
    Only stars within $600\,pc$ of the Sun and with a Galactic latitude larger than $60\degree$ are considered here (1,233,947 stars). We measured about 29,000 pairs in excess of accidental pairs.
    }
\label{pair-dist}
\end{center}
\end{figure}

Figure \ref{pair-dist} shows the angular separation distribution $\delta$ for the pairs of stars closer than $600\,pc$.
We first observed a clear excess of pairs for $\delta<20''$, in agreement with \citet{Zavada2020}, which demonstrates the existence of Gaia-detectable bound systems within $600\,pc$, in contrast to the situation in Fig. \ref{corr_HST} which showed the inability of HST to separate bound systems in the LMC.
Second, we noted that similar to HST, our algorithm failed to separate stars that are too close in angular distance, in this case around $0.5''$. The question of the separation limit, as it also relates to Gaia's scanning law, is beyond the scope of this paper (see \citet{these_blaineau} for more details). For our study, it is sufficient to know that we decided to focus only on pairs separated by at least $2''$ in order not to be limited by the resolution of Gaia. We have shown that the results of the next sections vary by less than $2\%$ (half of the estimated uncertainty) if we change this minimal separation in the range $[1.5''-2.5'']$ \citep{these_blaineau}.

\begin{figure}[htbp]
\begin{center}
    \includegraphics[width=9cm]{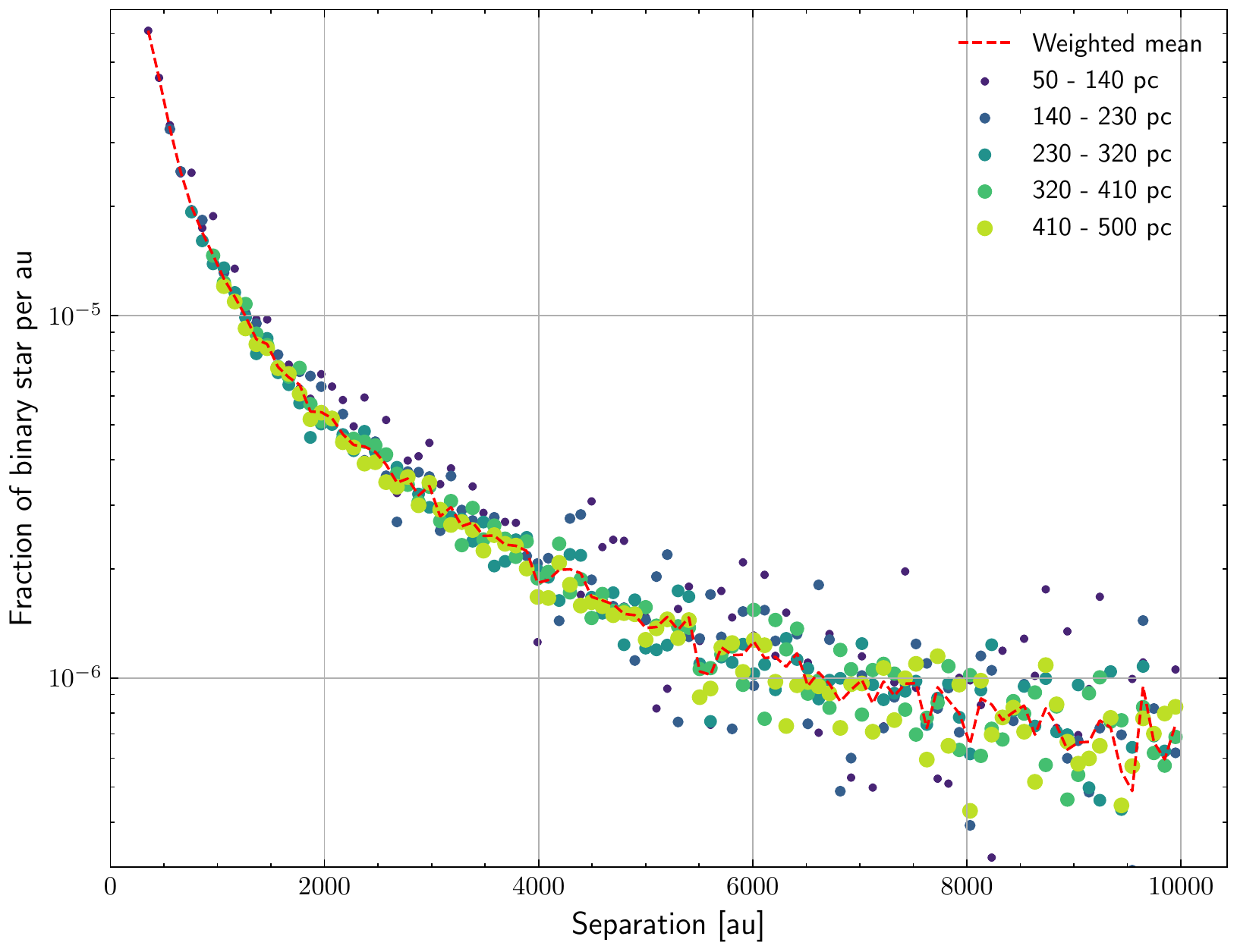}
    \caption[]{
    Differential rate of binaries $f_{bin}(a_t)$ measured in the five shells as a function of transverse separation $a_t$. The red dashed line gives the means for all shells. Error bars are not shown to avoid overloading the figure, but the scatter of each series gives an indication.
    }
\label{taux-mesure}
\end{center}
\end{figure}

To estimate the binarity rate of stars in a shell, we first built the distribution of the type shown in Figure \ref{pair-dist}. We then calculated the number of pairs remaining after subtracting the component due to accidental alignments, as a function of the separation. This expected random component is a linear function, fitted between
$60''$\footnote{Two stars at $\sim 450\,pc$ separated by $60''$ would be separated by only $0.5''$ if they were in the LMC. This directly illustrates the fact that an excess count of pairs in the LMC is only expected for separations much smaller than the resolution limit of HST.}
and $120''$ (dashed line in Figure \ref{pair-dist}).
Once the angular separation was converted to physical transverse separation $a_t$ (by using the measured parallax $\pi$), from the excess in each channel of $a_t$ we could derive
an estimate of the differential stellar binarity rate $f_{bin}(a_t)$ defined by expression (\ref{equation:fbin}), plotted in Figure \ref{taux-mesure} for each shell.
We would like to remind readers that this quantity represents the differential fraction of stars in pairs in excess of the expected number of accidental pairs separated by $a_t$, per unit of $a_t$.
The figure shows that the rates found do not depend on the shell considered, which allowed us to use the average distribution, weighted by the effective number of stars in each shell. These effective numbers are the numbers of stars found within the intervals of parallax of each shell, which were statistically corrected for misattributions of pairs
in the shells, due to uncertainties in the measured parallaxes
\footnote{
These uncertainties result in uncertainties on the distance of the stars; a pair can then be either misattributed if both components are effectively in another shell, or it can be missed if only one of the components is in another shell.
}
(see also Table \ref{tab:shellcounts}).
It should be noted that the closest shells are obviously those that allowed us to estimate the binarity rate at the smallest separations $a_t$, but at the expense of the statistics limited by the small volume of the shell. For shells at larger distances, the statistics is larger, but our angular separation limit of $2''$ prevents estimates at the smallest physical separations.

\section{Estimation of the local stellar binarity rate}
\label{section:fit}
In this section, we explain how we fit and extrapolated the differential binarity rate down to $50\,au$ transverse separations.
Following \citet{Duquennoy1991} and \citet{Raghavan2010}, we considered a log-normal model for the distribution of semi-major axes $a$ of binary systems. Since we measured projected separations $a_t$ and not semi-major axes, this distribution was modified as follows, by considering that the orbits have no preferred inclination:
\begin{equation}
\label{eq:proj_ln}
    f_{bin}^{DM}(a_t) = A\int^{+\infty}_{a_t} \frac{a_t}{r^2 \sigma \sqrt{2\pi(r^2-a_t^2)}} \exp{-\frac{\left(\ln{r/r_\text{mode}} -\sigma^2\right)^2}{2 \sigma^2}} d r,
\end{equation}
where $A$ and $\sigma$ were fitted to our data once $r_{mode}$ was chosen, corresponding to the maximum of the distribution.
\begin{figure}[htbp]
\begin{center}
 \includegraphics[width=9cm]{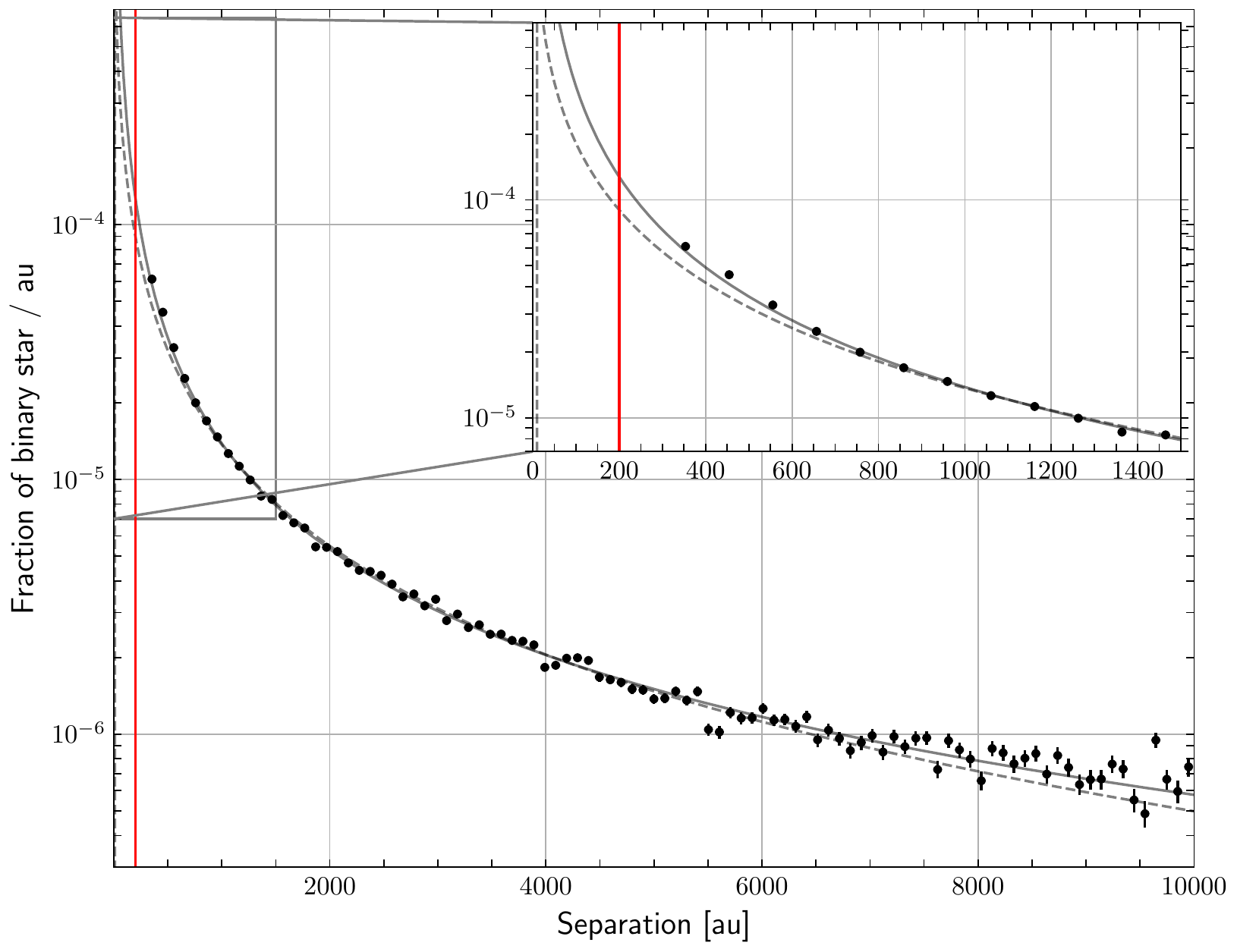}
    \caption[]{
        Projected log-normal function $f_{bin}^{DM}$ for $r_{mode} = 0.1\,au$ (dotted line), and power function $f_{bin}^{PL}$ (full line), fitted to the differential fraction of binary, as a function of the projected separation. The vertical red line corresponds to a projected separation of $200\,au$. The inset shows that the fit of $f_{bin}^{DM}$ is not satisfactory at small separations.
    }
\label{taux-fitte}
\end{center}
\end{figure}
If we chose $r_{mode}=0.1\,au$, then we find $A=0.126$ and $\sigma=2.72$,
but the fit is unsatisfactory for $a_t<600\,au$, and this is the case regardless of the parameter $r_{mode}$ (Figure \ref{taux-fitte}). This poor agreement can probably be attributed to the fact that this log-normal distribution was established for solar-type stars, while we studied a different and more extended population.
Nevertheless, following this fitted model, we computed the binarity rate for systems with $a_t>200\,au$
\footnote{This length corresponds to the most likely projected Einstein radius (at the LMC) of a $\sim 50M_{\odot}$ Galactic halo lens magnifying light from a LMC source.}
using Eqs (\ref{equation:bintot}), then (\ref{equation:proba}) and (\ref{equation:taux}), and found $F_{bin}(200\,au)=2.76\pm 0.03\%$.
By varying the $r_{mode}$ parameter within $10^{-4}\,au<r_{mode}<10^2 \,au$, we found values close to each other, contained within $2.05\%<F_{bin}(200\,au)<2.8\%$. Similarly, we found that $F_{bin}(100\,au)<3.6\%$
regardless of the $r_{mode}$ parameter.

We also fitted the following empirical power-law parameterization to our measurements:
\begin{equation}
f_{bin}^{PL}(a_t) = (0.19\,au^{-1})\times (a_t/1\,au)^{-1.379}, \label{eq:power_law}
\end{equation}
which better describes our data for small $a_t$ values
(Figure \ref{taux-fitte}). With this alternative model for the distribution of projected separations of binaries, we found that
$F_{bin}(200\,au)=3.48\pm 0.03\%$,
a value compatible with that found with the previous model.

In section \ref{section:method}, we mentioned that the total probability of a Gaia object being a member of a resolved binary system is less than the value given by expression (\ref{equation:bintot}). The value we found by integrating Eq. (\ref{eq:power_law}) from $a_t=50\,au$ is $0.114$, which is an upper limit of the fraction of Gaia objects belonging to binaries resolved in our sample. This number is indeed small compared to two, and its exact value does not impact
the computation of $F_{bin}$ from Eq. (\ref{equation:taux}),
especially since we are interested in the upper limit. It should be further noted that nothing can be said from our data about the binarity rate for $a_t<50\,au$ because extrapolation below this value is not constrained.

\section{Extrapolation at the LMC; impact of binarity on microlensing detection}
\label{section:impactLMC}
The previous study concerns a population of nearby Milky Way stars with $-0.5<G<9.5$. In the LMC, they would have an apparent magnitude $18<g<28$, that is they would be among the faintest stars in a classical catalog searching for gravitational microlensing effects.
This is a limitation of the extrapolation of this work to observations toward the LMC, in addition to the fact that we assume that this stellar population has the same binarity statistical characteristics in the LMC as in the Milky Way disk.

We saw in section \ref{section:blending} that if the angular separation of the components of a blend is much smaller than the angular Einstein radius -- expressed by $a_t << R_{E}/x$, where $x$ is given by Eq. (\ref{equation:x}) --, then both sources undergo roughly the same magnification and everything happens as if there were only one source. Ignoring this blending has therefore no impact on the detection efficiency of the ground surveys. As a consequence, only the pairs with $a_t \sim R_E/x$ or $\ge R_E/x$ should be considered for a possible impact on the statistics of microlensing effects.
From the $f_{bin}^{PL}(a_t)$ function of the differential binarity rate and using Eq. (\ref{equation:taux}), we can estimate the maximum proportion of situations where binarity induces a risk of failure to detect microlensing effects due to light-curve distortion.
\begin{figure}[htbp]
\begin{center}
   \includegraphics[width=9cm]{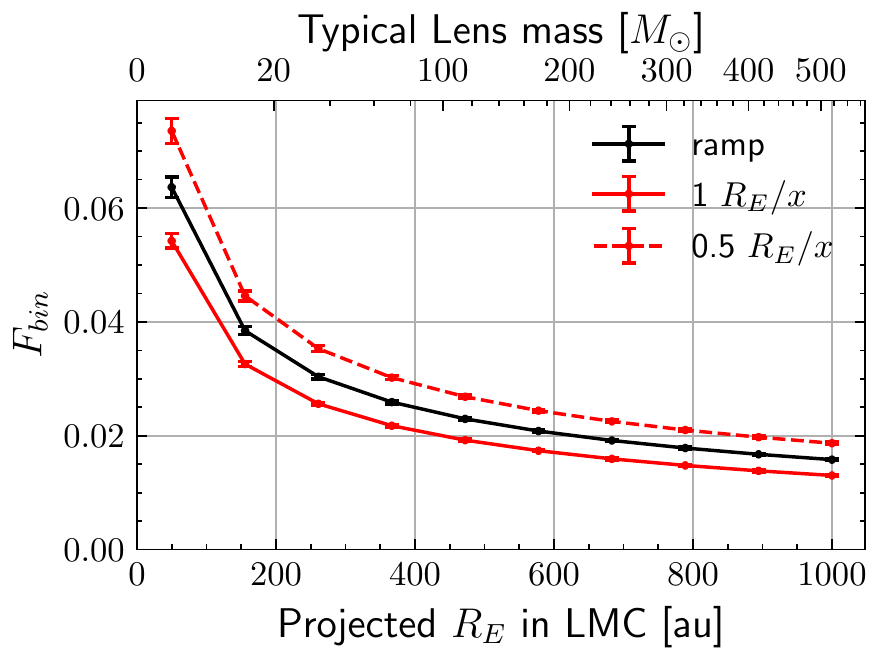}
    \caption[]{
Rates of binarity $F_{bin}$ as a function of the projected Einstein radius of the lens $R_E/x$, obtained for different ways of integration: from a component separation larger than $R_E/x$ (solid red line), or larger than $R_E/2x$ (dashed red line), or by weighting contributions with a ramp function (black line) (see text).
Typical masses corresponding to the Einstein radii are plotted on the upper abscissa for sources within the LMC.
This rate is an upper limit to the risk of not detecting microlensing effects due to binarity-induced distortion of the light-curve.
    }
\label{fraction-projete}
\end{center}
\end{figure}
Figure \ref{fraction-projete} shows this maximum proportion as a function of the Einstein radius projected in the LMC, $R_E/x$, after integration over the distribution of $a_t$ under three different assumptions: assuming that the blend effect is significant as soon as $a_t>R_E/2x$ (and thus integrating the differential distribution from $R_E/2x$ to infinity), or only when $a_t>R_E/x$, or weighting the differential distribution between zero and one (ramping) when $a_t$ varies from $0.1R_E/x$ to $1.75R_E/x$. The latter assumption was derived from the study of \citet{Griest1992}, which discusses the distortions expected by a microlensing curve for a composite source in detail.
Figure \ref{fraction-projete} shows, for example, that under the most pessimistic assumption (a blending effect taken into account as soon as $a_t>R_E/2x$), no more than $7\%$ of the sources are binary systems with a luminosity difference of less than 2.5 magnitude between the components, a configuration that could affect the identification of light-curves of microlensing events when $R_E/x>50\,au$.

\section{Discussion}
\label{section:discussion}
\subsection{Field of application}
One must first remember the limitations of this work. Only the population of stars of absolute magnitude $-0.5<g<9.5$ could be thoroughly studied, and any use for another population is an extrapolation, either for an identical population though in another galaxy possibly with a different metallicity, or for a stellar population in the Milky Way with more extensive types.
\begin{figure}[htbp]
\begin{center}
    \includegraphics[width=8cm]{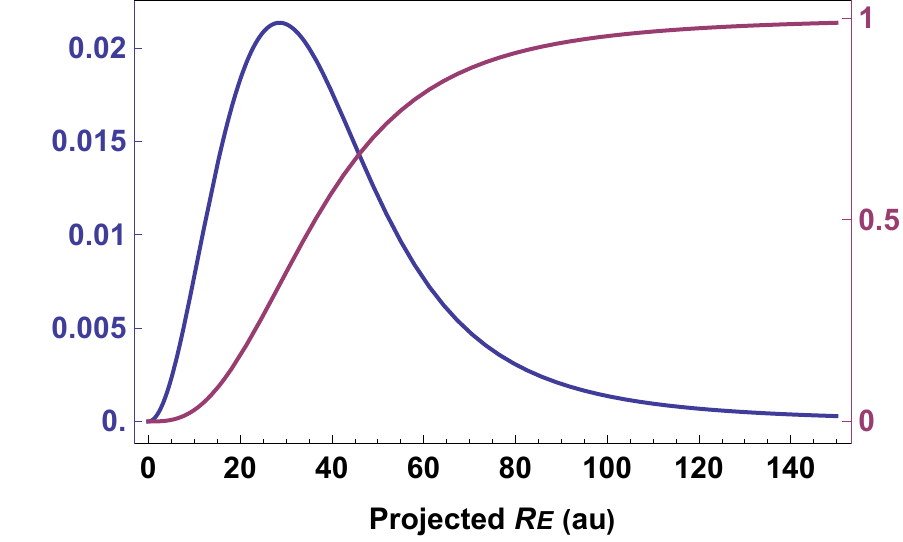}
    \caption[]{
    Distribution of the projected Einstein radius $R_E/x=R_ED_S/D_L$ for the microlensing events toward the LMC due to $1M_{\odot}$ lenses within the standard Galactic halo (blue line). The distribution expected for other lens masses $M_L$ can be obtained by scaling the abscissa by the factor $\sqrt{M_L/M_{\odot}}$. The purple line shows the cumulative distribution.
    The most probable value of $R_E/x$ is $28.6\,au$, the mean is $44.0\,au$, and the median is $36.5\,au$.
    }
\label{dist-REp}
\end{center}
\end{figure}

To compute the fraction of potentially complex events due to binarity as a function of the lens mass, we combined the rates from Fig. \ref{fraction-projete} with the projected Einstein radius $R_E/x$ distribution expected for that lens mass. Figure \ref{dist-REp} shows this generic distribution for lenses of $1M_{\odot}$ mass, changing with $M_L$ by simply scaling the abscissa with $\sqrt{M_L/M_{\odot}}$.
This has been established by assuming that the lenses are spatially distributed according to the standard Galactic halo described in \citet{EROSMACHOcombined}.
Table \ref{tab:resultats} shows, for a series of lens masses, the maximum expected fractions of situations where the shape of the microlensing events can be affected by binarity,
splitted into three domains of projected Einstein radius $R_E/x$. For each lens mass, the event fraction for each domain was deduced from the properly scaled distribution of Fig. \ref{dist-REp}: For $R_E/x<50\,au$, we conservatively assumed that $100\%$ of the events can be affected by blending due to binarity; for $50\,au<R_E/x<1000\,au$, we integrated the most pessimistic function of Fig. \ref{fraction-projete}, weighted by the normalized distribution of $R_E/x$ ; for $R_E/x>1000\,au$, we considered that no more than $2\%$ of the events can be affected by blending due to binarity.

\begin{table}
    \begin{center}
\begin{tabular}{lccccc} \hline \\  [-1ex]
$M_L$ ($M_{\odot}$) & 1 & 3 & 10 & 30 & 100 \\
\hline
$R_E/x<50\,au$ & 71\% & 34\% & 9\% & 2\% & 0.4\% \\
max. contribution & 71\% & 34\% & 9\% & 2\% & 0.4\% \\
\hline
$50<R_E/x<1000$ & 29\% & 66\% & 91\% & 97\% & 95\%  \\
max. contribution & 1.8\% & 3.9\% & 4.6\% & 4.0\% & 4.8\% \\
\hline
$R_E/x>1000\,au$ & 0\% & 0\% & 0.2\% & 0.8\% & 5\% \\
max. contribution & 0\% & 0\% & 0\% & 0\% & 0.1\% \\
\hline
Total & 73\% & 38\% & 14\% & 6.2\% & 5.3\% \\
\hline
\end{tabular}
\centering
\caption[]{Fractions of events toward the LMC within the three domains of the projected Einstein radius and their {\it maximum} contributions to the fraction of events with significant blending due to binarity.
}
\label{tab:resultats}
\end{center}
\end{table}

The totals given in the table are the maximum proportions of events
whose shape can be affected by binarity. This does not necessarily mean that such events systematically escape detection (a lot depends on the selection algorithm).
Nevertheless, for the sake of caution, we will keep these proportions as an upper limit for situations where the classic detection efficiency calculation does not apply.
In the absence of a specific simulation, the status of these events on binaries in terms of the detection efficiency is indeed poorly known, and
this must be taken into account as a systematic uncertainty for the measurement of optical depths and event rates.

Our study shows that the impact of source binarity can be neglected to first order when searching for gravitational microlensing effects from lenses heavier than $30M_{\odot}$.
The maximum value of uncertainty of $6.2\%$ is even smaller when the lenses are more massive (Fig. \ref{dist-REp}), showing that the additional effect of binarity on the blending effects
can be neglected for heavy lenses, as was done in \citet{EROSMACHOcombined}.
Figures \ref{fraction-projete} and \ref{dist-REp} allows for these numbers to be estimated in the case of even more massive lenses toward the LMC.
For an estimate corresponding to another Galactic halo model, or to other targets than LMC, Fig. \ref{dist-REp} needs to be rebuilt.

For events due to lighter lenses with a higher probability of projected Einstein radii $R_E/x<50\,au$, it is currently not possible to draw reliable conclusions on the impact of the blend due to the binarity of the LMC sources with our technique without risky extrapolation. An alternative method for estimating the differential binarity rate with a projected separation of less than $50\,au$ is needed, coupled with a specific simulation to estimate the detection efficiency of non-PSPL events.

\citet{Skowron} found that $0.5\%$ of the events observed by OGLE toward the Galactic Center (not LMC) have two peaks because they are due to binary systems (lens or source). This suggests that a larger proportion must suffer lesser distortions  \citep{Distefano1996}, but this leads in particular to them being assigned an incorrect duration, and this may affect the efficiency calculation.
It is not easy to extrapolate these observations to the LMC
(but see \citet{Griest1992}),
but nonetheless it seems to us that the question is not strictly closed for microlensing toward the LMC caused by light halo objects, even if it is likely that the effect we are talking about is negligible.

\subsection{Discussion on stellar populations}
Since the majority of the stars observed in the LMC correspond to the brightest stars in the Gaia data that we used (Fig. \ref{mag-vs-dist}), we investigated the variation of the binarity rate $F_{bin}(200\,au)$ with the magnitude of the stars. We therefore reproduced our analysis by restricting our sample to two absolute magnitude $g$ domains, between -0.5 and 4.5 mag ($3,090,000$ stars) and between 4.5 and 9.5 mag ($612,000$ stars), always limiting the maximum difference in magnitude between the two components to 2.5. The statistical power here was significantly reduced for the population with $-0.5<g<4.5$,
but we nevertheless found a significantly lower binarity rate for bright star pairs ($F_{bin}(200\,au) = 1.42\pm 0.08\%$) than for faint star pairs ($F_{bin}(200\,au) = 2.58\pm 0.04\%$). It is therefore likely that in LMC catalogs such as the EROS2 or MACHO surveys, which are composed of rather bright stars, we overestimated the binarity rate by assuming that the stellar populations from our Gaia sample and in the LMC are similar.

\begin{figure}
    \centering
    \begin{subfigure}{0.5\linewidth}
        \includegraphics[width=\textwidth]{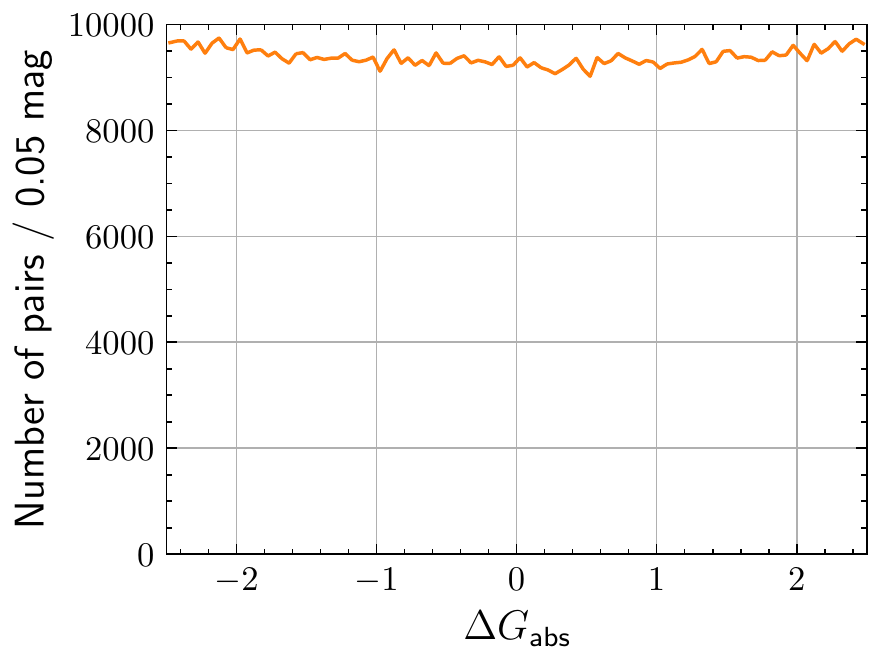}
        \caption{Accidental pairs}
        \label{fig:dmag_random}
    \end{subfigure}    
    \begin{subfigure}{0.5\linewidth}
        \includegraphics[width=\textwidth]{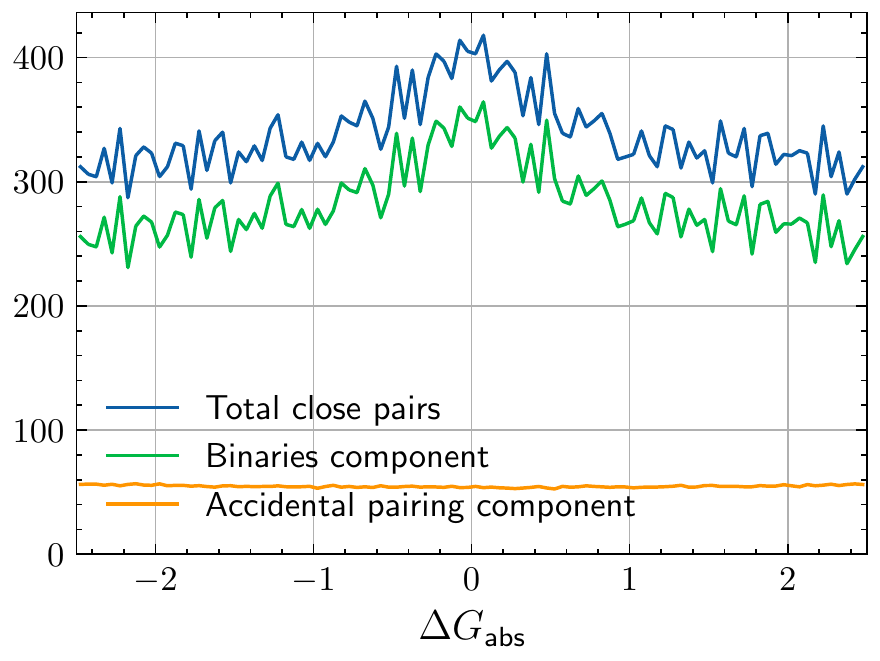}
        \caption{Pairs $1000\,au < a_t < 2500\,au$}
        \label{fig:dmag_2500}
    \end{subfigure}\hfill
    \begin{subfigure}{0.5\linewidth}
        \includegraphics[width=\textwidth]{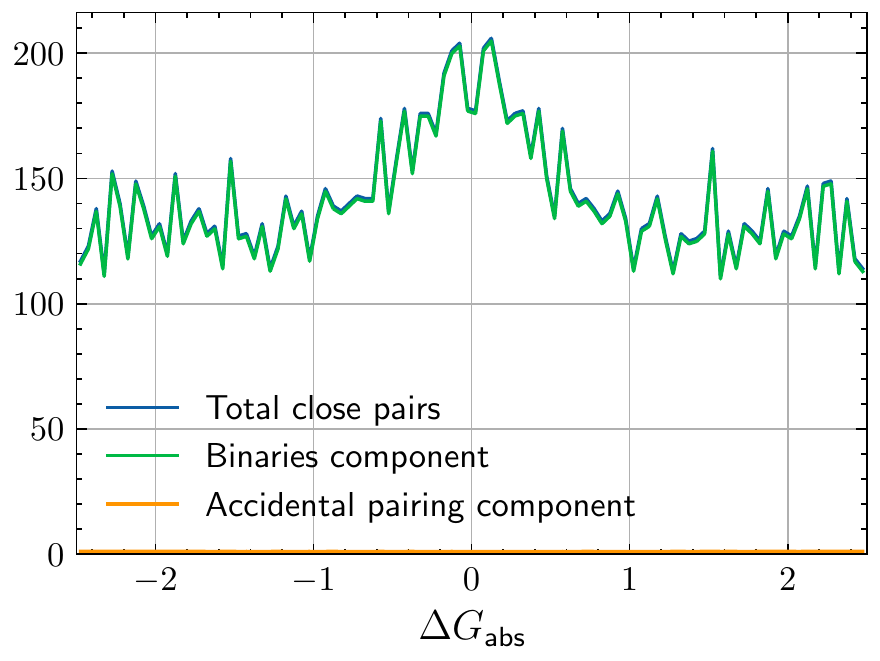}
        \caption{Pairs  $a_t < 1000\,au$}
        \label{fig:dmag_1000}
    \end{subfigure}
    \caption{
    Magnitude differences between components of pairs, for different separation ranges.
    (a) Magnitude differences for pairs separated by more than $30000\,au$ (only due to fortuitous alignments).
   (b) and (c)
Same distributions for pairs separated by $1000<a_t<2500\,au$ and $a_t<1000\,au$ (blue), distributions for the fortuitous alignments (orange), normalized according to the sample size, and the difference between the two distributions (green; {i.e.,} the distribution of the magnitude difference for binary systems only). 
    These histograms are sampled in 0.05 magnitude steps with one entry per star.
    In panel 
    (c) the number of fortuitous alignments is negligible, and thus the orange distribution is almost zero and green and blue distributions are superimposed.
    }
    \label{diff-mag}
\end{figure}
We also investigated whether there are correlations between the magnitude differences and the separation $a_t$ of binaries.
Figure \ref{diff-mag}
shows in green the distributions of magnitude differences $\Delta G$ between components for binaries with $1000\,au<a_t<2500\,au$ (Fig. \ref{diff-mag}(b)) and with $a_t<1000\,au$ (Fig. \ref{diff-mag}(c)), obtained by subtracting the expected distributions for random pairs (Fig. \ref{diff-mag}(a)), properly normalized, from the observed (blue) distributions.
The distribution for random pairs was deduced from that of the difference $\Delta G$ of pairs of uncorrelated stars, separated by $a_t>30000\,au$
-- which turns out to be approximately uniform.
It appears that the closer the binary system is, the smaller the difference in magnitude between components. It is tempting to explain this difference by the intervention of a gravitational capture mechanism, which would favor the formation of distant binaries, whose luminosity would be consequently less correlated. These observations are corroborated by the distribution of mass ratios as a function of the semi-major axis or period of the binary systems in \citet{2017Moe_DiStefano}.

Finally, we examined the case of red giants, which constitute an important part of the catalogs of the historical microlensing surveys to the LMC. Unfortunately, they represent too few stars in our Gaia sample to establish a reliable binarity rate. However, we examined the magnitude distribution of stars in pairs with separation $a_t<2500\,au$ containing at least one giant. In this sample of pairs, we found that there are almost no giant binary systems (less than $1\%$ of the sample), but only binaries consisting of a giant and a main sequence star, with magnitude difference $\Delta G$ larger than $2.5$ in $85\%$ of the cases. This fact reinforces our conclusion that we probably overestimated the proportion of binaries in EROS2/MACHO-type catalogs by extrapolating the binarity rate measured in Section \ref{section:impactLMC}.

\section{Conclusions: The impact of binarity on microlensing surveys}
\label{section:conclusion}
We conclude from this study that the search detection efficiency of long duration microlensing events due to lenses heavier than $30M_{\odot}$ toward the LMC is not significantly affected by the source binarity.
This result is useful not only in the recent combined analysis of EROS and MACHO data spanning several decades \citep{EROSMACHOcombined}, but also in future research with Rubin-LSST.

On the other hand, for lenses lighter than $30M_{\odot}$, as soon as the projected Einstein radius is less than a few tens of $au$, the binarity rates extrapolated here are higher and less reliable. The fraction of events that may be affected by blending due to source binarity becomes less negligible, and another study must be undertaken to estimate the impact of a binarity rate that may be more significant and probably positive on the detection efficiency.

Although it is not possible to estimate from the Gaia database the differential binarity rate for $a_t<50\,au$ in the LMC, one can conversely consider detecting binarity effects by measuring distortions due to blending with respect to a simple (PSPL) microlensing effect. The fraction of cases where a fit with blend (according to Eq. (\ref{eq:blend_flux})) is significantly better than the PSPL fit would allow us to extract constraints on the binarity rate. This should be particularly relevant to Rubin-LSST observations if the photometric accuracy reaches a few milli-magnitudes.

Our study of the impact of binarity rates on microlensing detection efficiency toward the LMC is easily transferable, through some scaling and modelization of the lens spatial distribution, to studies of microlensing within the Galactic plane. In this case, the source population should better resemble the one studied in this paper.
Our last comment is that the use of a tolerant prefiltering, not too sensitive to the precise shape of the magnification curve, remains the safest technique to mitigate the possible effects of distortion due to the binarity of the source on the detection efficiency.

\begin{acknowledgements}
We thank Olivier Perdereau for his useful comments on the manuscript. This work was supported by the Paris Île-de-France Region.
\end{acknowledgements}
\bibliographystyle{aa}
\bibliography{citations_parallax}
\end{document}